\documentclass[conference,10pt]{IEEEtran}
\usepackage{epsfig,rotating,setspace,latexsym,amsmath,epsf,amssymb,amsfonts,bm,theorem,subfigure,epstopdf}

\usepackage{amsbsy}
\usepackage{cite,authblk}
\usepackage{bbm}
\usepackage{color, xcolor}
\usepackage{mathtools}
\usepackage{algorithm}
\usepackage{textcomp}
\usepackage[noend]{algpseudocode}
\usepackage{verbatim}
\usepackage{graphicx, graphics}
\usepackage{afterpage}
\usepackage{placeins}
\usepackage{bm}
\usepackage{authblk}
\PassOptionsToPackage{bottom}{footmisc}
\PassOptionsToPackage{symbol}{footmisc}
\usepackage{footmisc}
\usepackage{multirow}
\usepackage{booktabs}
\usepackage{float}
\usepackage{afterpage}

\DeclareMathOperator*{\argmax}{arg\,max}

\algrenewcommand\algorithmicforall{\textbf{foreach}}
\algrenewcommand\algorithmicindent{.8em}

\IEEEoverridecommandlockouts
\allowdisplaybreaks

\usepackage{graphicx}
\usepackage{epsfig}
\usepackage{epstopdf}
\begin{document}

\title{Transformer-Driven Neural Beamforming with Imperfect CSI in Urban Macro Wireless Channels }

\author{
Cemil Vahapoglu$^{1,2}$, Timothy J. O’Shea$^{2}$, Wan Liu$^{2}$, Tamoghna Roy$^{3*}$\thanks{$^*$This work was conducted while the author was with Deepsig Inc.}, Sennur Ulukus$^{1}$ \\
\normalsize $^{1}$University of Maryland, College Park, MD, $^{2}$DeepSig Inc., Arlington, VA, $^{3}$Meta Reality Labs, New York, NY \\
\normalsize \emph{cemilnv@umd.edu, tim@deepsig.ai, wliu@deepsig.ai, tamoghna.roy@gmail.com, ulukus@umd.edu}}

\maketitle

\begin{abstract}
The literature is abundant with methodologies focusing on using transformer architectures due to their prominence in wireless signal processing and their capability to capture long-range dependencies via attention mechanisms. In particular, depthwise separable convolutions enhance parameter efficiency for the process of high-dimensional data characteristics of MIMO systems. In this work, we introduce a novel unsupervised deep learning framework that integrates depthwise separable convolutions and transformers to generate beamforming weights under imperfect channel state information (CSI) for a multi-user single-input multiple-output (MU-SIMO) system in dense urban environments. The primary goal is to enhance throughput by maximizing sum-rate while ensuring reliable communication. Spectral efficiency and block error rate (BLER) are considered as performance metrics. Experiments are carried out under various conditions to compare the performance of the proposed NNBF framework against baseline methods zero-forcing beamforming (ZFBF) and minimum mean square error (MMSE) beamforming. Experimental results demonstrate the superiority of the proposed framework over the baseline techniques.
\end{abstract}

\section{Introduction}
Conventional wireless communication systems are significantly dependent on probabilistic modeling and signal processing methods. Nevertheless, their efficacy is constrained by computing complexity and inconsistencies between theoretical models and the the real-time implementation \cite{erpek2020deep}. Numerous approaches presume static network conditions, resulting in poorer performance in dynamic contexts due to the reliance on algorithm initialization \cite{2011iterativeWeightedMMSE, 2008_WSR_WMMSE}. In contrast, machine learning (ML) allows systems to adjust by acquiring knowledge from dynamic spectrum data instead of relying on predetermined policies \cite{ClancyCognitive}. Improvements in high-performance computing (HPC) and data accessibility have significantly augmented deep learning methodologies, equipping them with robust representational capacities \cite{AppML}. Using deep neural networks (DNNs) as universal function approximators, they can proficiently tackle complex resource management challenges, such as beamforming and power control, by learning non-linear mappings often managed by conventional algorithms \cite{AppML, Sun2018}.

Depthwise separable convolutions enhance parameter efficiency in extracting spatial and temporal features. They reduce computational complexity without sacrificing the performance of convolutional neural networks (CNNs) \cite{Guo2018NetworkDF}, making them valuable in resource-constrained scenarios such as mobile edge computing, dynamic beamforming adaptation, and high-dimensional MIMO data processing \cite{Naeem2021}. In addition, transformers have become powerful models for wireless signal processing, capturing long-range dependencies through attention mechanisms. Unlike CNNs, which rely on local receptive fields, transformers aggregate channel state information (CSI) and dynamic spectral data globally, enabling more adaptive and data-driven beamforming solutions. Recent studies demonstrate that attention-based models outperform conventional deep learning models in capturing channel dependencies and optimizing power allocation under dynamic network conditions \cite{DT_transformer,atteion_based_nn}.

In this work, we propose a novel deep learning (DL) framework, which consists of depthwise separable convolutions and transformers, for uplink multi-user single-input multiple-output (MU-SIMO) beamforming design in dense urban environments, which are modeled by urban macro (UMa) channel. The UMa channel provides a realistic representation of urban wireless environments by capturing the key characteristics of dense urban deployments \cite{3gppTR38901}. The proposed framework is denoted as NNBF for the remainder of the paper. NNBF performs unsupervised training, targeting sum-rate maximization across single-antenna user equipment (UE). In our previous works, we proposed simpler deep learning frameworks for sum-rate maximization in uplink MU-SIMO and joint power allocation $\&$ beamforming design in downlink multi-user multiple-input single-output (MU-MISO) settings for TDL channel models \cite{vahapoglu2023deep, vahapoglu2024_joint_power_bf}. However, it is assumed that the perfect channel state information (CSI) is available for these works. In this work, it is considered that only imperfect CSI, which is obtained by channel estimation in the radio unit (RU), is available. We perform the performance analysis of our proposed framework by comparing with zero-forcing beamforming (ZFBF) and minimum mean square error (MMSE) beamforming, which are considered as our baselines. Spectral efficiency and block error rate (BLER) are considered as performance metrics. Experimental results for diverse settings demonstrate the superiority of the proposed framework NNBF compared to baseline techniques ZFBF and MMSE.

\section{System Model \& Problem Formulation}

\subsection{Uplink Multi-User SIMO (MU-SIMO) Setup}
\begin{figure}[t]
    \centerline{\includegraphics[width= 1\linewidth]{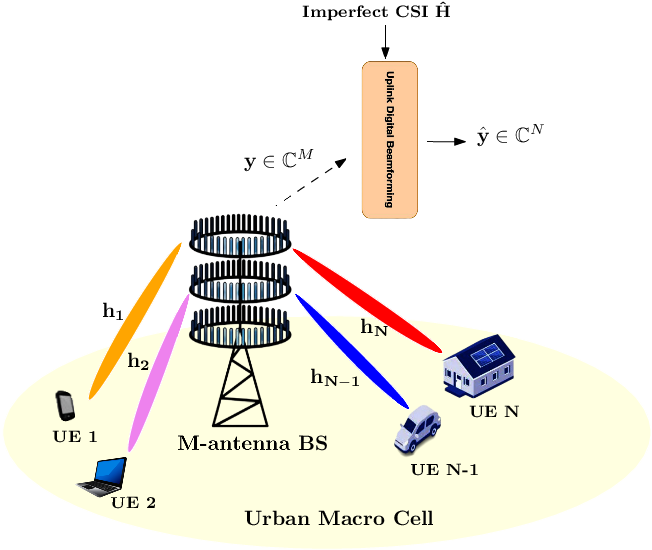}}
    \caption{Uplink multi-user SIMO system in a dense urban environment, where single-antenna UEs transmit data streams on the same time/frequency resources and M-antenna BS apply digital beamforming on the received signal $\mathbf{y}$.}
    \label{System Model}
\end{figure}

We consider an uplink transmission scenario in which $N$ single antenna UEs send data streams to a base station (BS) equipped with $M$ receive antennas as shown in Fig.~\ref{System Model}.

The uplink channel matrix $\mathbf{H} = 
[\mathbf{h}_1  ~ \mathbf{h}_2 ~ \cdots ~ \mathbf{h}_N] \in \mathbb{C}^{M \times N}$, where $\mathbf{h}_k$ denotes the channel vector between UE $k$ and the BS. The received signal $\mathbf{y}$ can be expressed as
\begin{align} \label{received_signal}
    \mathbf{y} = \sum_{i=1}^N \mathbf{h}_i x_i +\mathbf{n} 
\end{align}
where $\mathbf{x} = [ x_1^H ~ x_2^H ~ \cdots ~ x_N^H]^H \in \mathbb{C}^N$ represents the transmitted signal, with each entry corresponding to the modulated data symbols sent by the UEs, satisfying $\mathbb{E}[x_k^Hx_k] =1, \, \forall k=1,\ldots, N.$ Additionally, $\mathbf{n} = [ n_1^H ~ n_2^H ~ \cdots ~ n_M^H]^H \in \mathbb{C}^M$ denotes the additive white Gaussian noise with i.i.d.~entries $n_l \sim \mathcal{CN}(0, \sigma^2), \, \forall l=1,\ldots, M$.

\begin{figure}[t]
    \centerline{\includegraphics[width= 1\linewidth]{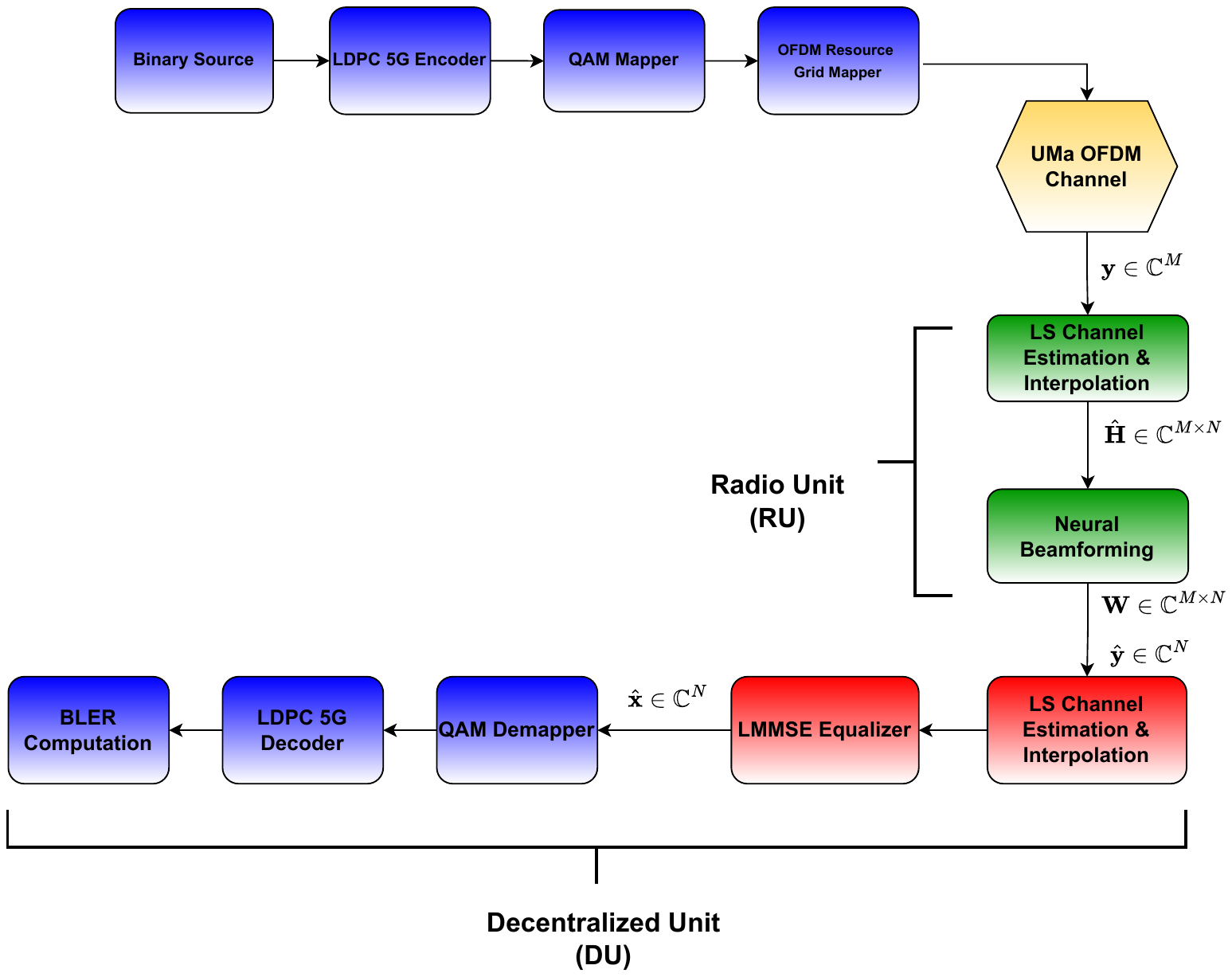}}
    \caption{The block diagram of system model.}
    \label{System Model Block Diagram}
\end{figure}

It is assumed that Uplink Performance Improvement (ULPI) specifications are adopted, particularly employing the Cat-B ULPI-B front-haul (FH) option. In this configuration, uplink channel estimation and beamforming are executed within the RU, while the distributed unit (DU) is responsible for both uplink channel estimation and uplink equalization. Therefore, the uplink channel estimate $\mathbf{\hat{H}} = 
[\mathbf{\hat{h}}_1  ~ \mathbf{\hat{h}}_2 ~ \cdots ~ \mathbf{\hat{h}}_N] \in \mathbb{C}^{M \times N}$ is calculated to facilitate beamforming design directly within the RU, ensuring that all the necessary uplink processing tasks for beamforming, including the beamforming design itself, are efficiently managed locally within the RU. The block diagram of the system model, illustrating the RU-DU split, can be seen in Fig.~\ref{System Model Block Diagram}.

Then, the received signal in (\ref{received_signal}) is processed using beamforming weights $\mathbf{W} = [\mathbf{w}_1  ~ \mathbf{w}_2 ~ \cdots ~ \mathbf{w}_N ] \in \mathbb{C}^{M \times N}$ to retrieve data symbols while power consumption of beamforming weights are checked to satisfy $\mathbf{w_k}^H\mathbf{w_k} \leq 1, \forall k=1, \ldots, N$. Specifically, $\mathbf{w}_k \in \mathbb{C}^{M}$ serves as the linear beamforming filter to estimate the transmitted data symbol of UE $k$, aiming to maximize throughput while mitigating the interference from other users
\begin{align}\label{received_uek}
    \mathbf{w}_k^T \mathbf{y} & = \sum_{i=1}^N \mathbf{w}_k^T\mathbf{h}_i x_i +\mathbf{w}_k^T\mathbf{n}
\end{align}

\subsection{Beamforming Design for Sum-rate Maximization Problem}
Our objective is to design beamforming weights that maximize the overall sum-rate across all UEs. The received signal for UE $k$ after beamforming with $\mathbf{w_k}$ can be reformulated as 
\begin{align} \label{received_uek_v2}
    \hat{y}_k &= \mathbf{w}_k^T \mathbf{y} \nonumber \\ 
     &= \underbrace{\mathbf{w}_k^T\mathbf{h}_k x_k}_{desired \, signal} + \underbrace{ \sum_{i=1, i\neq k}^N \mathbf{w}_k^T\mathbf{h}_i x_i}_{interfering \, signal} +\underbrace{\mathbf{w}_k^T\mathbf{n}}_{noise}
\end{align}

The first term in (\ref{received_uek_v2}) represents the desired symbol for UE $k$, while the second and the third terms denote the inter-symbol interference (ISI) from other UEs and receiver noise, respectively. Then, the signal-to-interference-plus-noise ratio (SINR) for UE $k$ is given by
\begin{align}\label{sinr}
    \gamma_k = \frac{|\mathbf{w}_k^T \mathbf{h}_k|^2}{\sum_{i=1, i\neq k}^N |\mathbf{w}_k^T\mathbf{h}_i|^2 + \mathbb{E}|\mathbf{w}_k^T \mathbf{n}|^2}
\end{align}

Thus, the beamforming design for sum-rate maximization can be formulated as 
\begin{align}\label{sum-rate maximization problem}
    \mathbf{W}^* = \argmax_{\mathbf{W}} & \quad \sum_{i=1}^N \alpha_i \log(1 + \gamma_i) \nonumber \\
    \textrm{s.t.} &\quad \textrm{tr}(\mathbf{W}^H\mathbf{W}) \leq N
\end{align}
where $\alpha_i$ denotes the rate weighting factor for each UE $i$, which is also a trainable parameter within our framework, ensuring that  $\sum_{i=1}^N \alpha_i = 1$.

\section{Proposed Deep Neural Network}
In this section, we introduce a DNN architecture to design beamforming weights for the sum-rate maximization problem outlined in (\ref{sum-rate maximization problem}). The network takes the IQ symbols of imperfect frequency domain channel response $\hat{\mathbf{H}}$ as input and outputs the beamforming weights $\mathbf{W}$ as defined in the system model. Throughout the rest of the paper, $B$ represents the batch size of the dataset while $L$ and $K$ indicate number of OFDM symbols and subcarriers respectively.

\subsection{Deep Neural Network (DNN) Architecture}
Proposed DNN architecture consists of two primary components: Convolutional Residual Network and Stacked Multi-Channel Attention as shown in Fig.~\ref{DNN_architecture}.

\subsubsection{Convolutional Residual Network}
The first part of the model employs a convolutional residual network, which combines regular and grouped convolutions \cite{Chollet2016XceptionDL}. At this stage, the network receives the imperfect frequency channel response $\hat{\mathbf{H}}$ as input, reshaped to dimensions $(B,2MN,L,K)$, is first applied padding along $(L,K)$ with mirror reflections of edge values.

\begin{figure*}[t]
    \centerline{\includegraphics[width=1.2\linewidth]{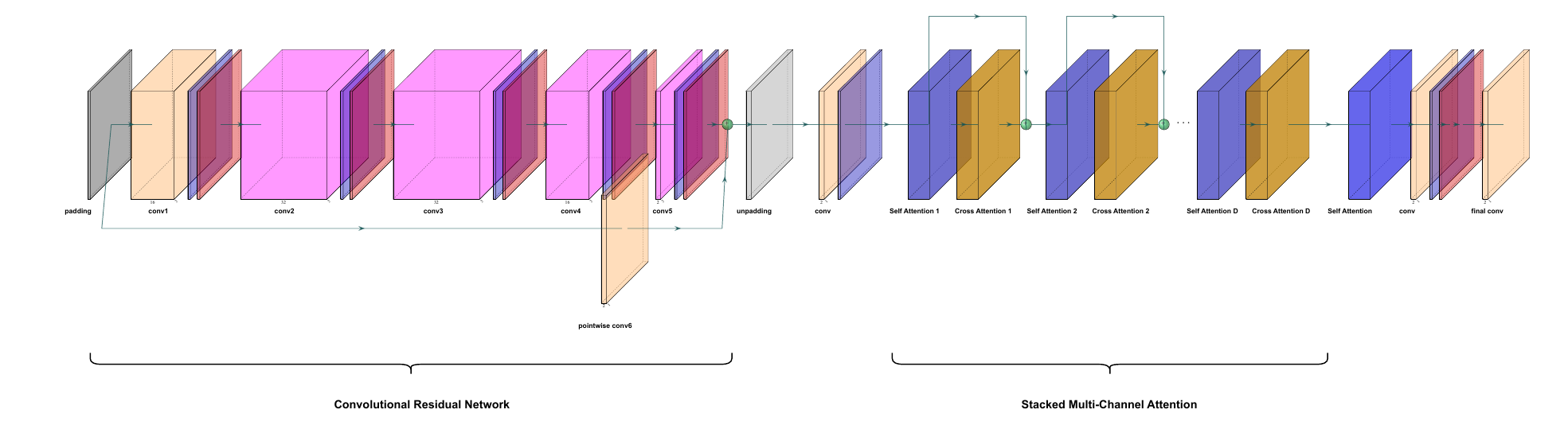}}
    \caption{Deep neural network architecture.}
    \label{DNN_architecture}
\end{figure*}

This is followed by a regular convolution (orange ``conv1'' in Fig.~\ref{DNN_architecture}) and a sequence of grouped convolutions (vivid magenta layers in Fig.~\ref{DNN_architecture}). It enables the capture of local features effectively while enhancing efficiency through the grouped convolutions \cite{Chollet2016XceptionDL}. Number of groups for grouped convolutions are determined as the minimum of the number of input channels and the number of output channels. Moreover, residual connection is integrated to facilitate the gradient flow while pointwise convolution follows residual connection to consolidate the features from the residual path. All convolutional layers except pointwise convolution are followed by batch normalization and GELU activation function.

\subsubsection{Stacked Multi-Channel Attention}
The second part of the model utilizes self attention and cross attention mechanisms to capture intra-channel relationships and the inter-channel dependencies. In \cite{multi-channel-transformer-sr}, it was shown that combining self attention and cross attention mechanisms can effectively capture contextual relationships within and between channels for speech recognition. Our proposed architecture demonstrates that self attention and cross attention mechanisms are also valuable for interference mitigation in 5G MIMO networks, specifically in dense urban environments.

\begin{figure}[h]
     \centerline{\includegraphics[width=1\linewidth]{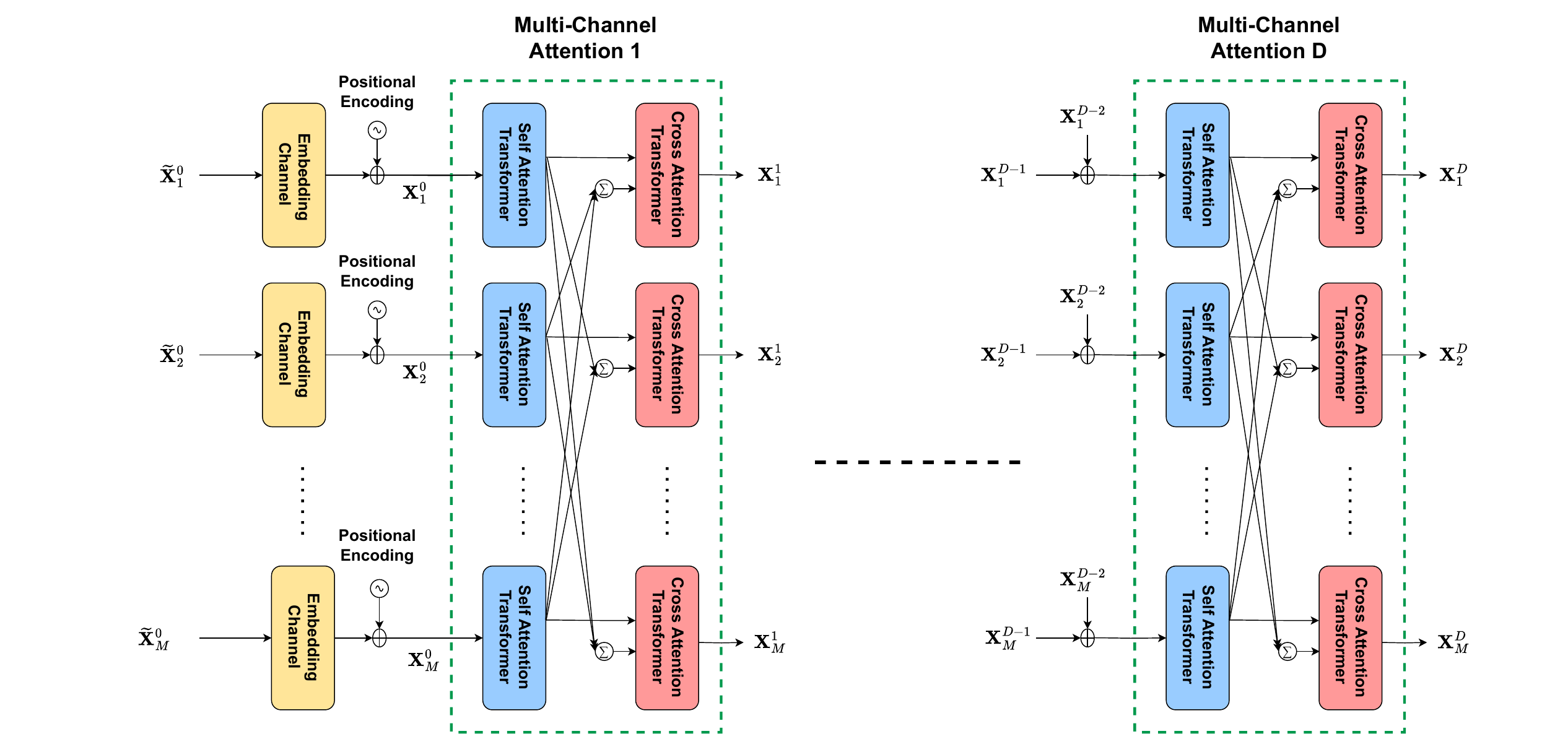}}
      \caption{Stacked multi-channel attention module.}
      \label{multi_channel_attention}
\end{figure}

Stacked multi-channel attention can be seen in Fig.~\ref{multi_channel_attention}. It is the repeated sequence of self attention and cross attention transformer layers. Before transformer layers, input features are divided into $M$ chunks along the channel dimension, i.e., input of shape $(B,\mathrm{depth}, L, K)$ is split into $M$ non-overlapping components, each with dimensions $(B,\frac{\mathrm{depth}}{M}, L, K)$. Divided features are projected to dense embedding spaces. Channel embeddings are $1\times1$ unbiased convolutional layers followed by batch normalization. Positional encoding is employed on embedded space representations. 

In self attention transformer layer, there are two sublayers as suggested by \cite{Vaswani2017AttentionIA}. First sublayer consists of scaled dot-product attention to compute the attention weights across time and frequency resources $(L,K)$ to score over channel dimensions. Queries, keys and values are computed by $1\times1$ convolution followed by GELU activation function. Specifically, for the $i$th embedding representation and $d$th self attention, they can be expressed as
\begin{align}
    \mathbf{Q}^{d}_{i,\textrm{sa}} &= \mathrm{GELU}\left(\mathbf{X}^{d-1}_i \circledast \mathbf{F}^{d}_{i,q}+ \mathbf{b}^{d}_{i,q} \right) \nonumber \\
    \mathbf{K}^{d}_{i,\textrm{sa}} &= \mathrm{GELU}\left(\mathbf{X}^{d-1}_i \circledast \mathbf{F}^{d}_{i,k}+ \mathbf{b}^{d}_{i,k} \right) \nonumber \\
    \mathbf{V}^{d}_{i,\textrm{sa}} &= \mathrm{GELU}\left(\mathbf{X}^{d-1}_i \circledast \mathbf{F}^{d}_{i,v}+ \mathbf{b}^{d}_{i,v} \right)
\end{align}
where $\mathbf{F}^{d}_{i,q}$, $\mathbf{F}^{d}_{i,k}$, $\mathbf{F}^{d}_{i,v}$ $\in \mathbb{R}^{\mathrm{dim}\times \mathrm{dim}}$ represent trainable convolutional filters while  $\mathbf{b}^{d}_{i,q}$, $\mathbf{b}^{d}_{i,k}$, $\mathbf{b}^{d}_{i,v}$ $\in \mathbb{R}^{\mathrm{dim}}$ are trainable bias parameters of $d$th self attention for input $\mathbf{X}_i^{d-1}$. $\mathrm{dim}$ corresponds to number of input and output channels $\frac{\mathrm{depth}}{M}$. Then, corresponding self attention output is computed as
\begin{align}
    \mathbf{A}^{d}_{i,\textrm{sa}} & = \mathrm{Softmax}\left( \frac{\mathbf{Q}^{d}_{i,\textrm{sa}} \left(\mathbf{K}^{d}_{i,\textrm{sa}}\right)^T}{\sqrt{\mathrm{dim}}}\right) \mathbf{V}^{d}_{i,\textrm{sa}}
\end{align}
In the second sublayer of self attention transformer, fully connected layers with a GELU activation function serve as the feed forward network, producing the final output of self attention transformer $\mathbf{Z}_i^d$. Channel dimension is fed into the fully connected layers as features, i.e., input of second sublayer is reshaped as $\left( BLK, \frac{\mathrm{depth}}{M} \right)$. Additionally, residual connections and normalization are used for each sublayer as described in \cite{Vaswani2017AttentionIA}.

In cross attention transformer layer, there are two sublayers similar to those in self attention. First sublayer consists of scaled dot-product attention weights across time and frequency resources $(L,K)$. Unlike self attention, keys and values are derived by performing $1\times1$ convolution on a weighted summation of other self attention outputs $\Bar{\mathbf{Z}}_i^d =  \sum_{j=1, j\neq i }^M \beta_j^d \mathbf{Z}_j^d$ where $\beta$ parameters are trainable as well. The queries are computed through $1\times1$ convolution on the self attention output of the same input $\mathbf{Z}^{d}_i $.
\begin{align}
    \mathbf{Q}^{d}_{i,\textrm{ca}} &= \mathrm{GELU}\left(\mathbf{Z}^{d}_i \circledast \tilde{\mathbf{F}}^{d}_{i,q}+ \tilde{\mathbf{b}}^{d}_{i,q} \right) \nonumber \\
    \mathbf{K}^{d}_{i,\textrm{ca}} &= \mathrm{GELU}\left(\Bar{\mathbf{Z}}_i^d\circledast \tilde{\mathbf{F}}^{d}_{i,k}+ \tilde{\mathbf{b}}^{d}_{i,k} \right) \nonumber \\
    \mathbf{V}^{d}_{i,\textrm{ca}} &= \mathrm{GELU}\left(\Bar{\mathbf{Z}}_i^d \circledast \tilde{\mathbf{F}}^{d}_{i,v}+ \tilde{\mathbf{b}}^{d}_{i,v} \right) 
\end{align}
where $\tilde{\mathbf{F}}^{d}_{i,q}$, $\tilde{\mathbf{F}}^{d}_{i,k}$, $\tilde{\mathbf{F}}^{d}_{i,v}$ $\in \mathbb{R}^{\mathrm{dim}\times \mathrm{dim}}$ represent trainable convolutional filters and  $\tilde{\mathbf{b}}^{d}_{i,q}$, $\tilde{\mathbf{b}}^{d}_{i,k}$, $\tilde{\mathbf{b}}^{d}_{i,v}$ $\in \mathbb{R}^{\mathrm{dim}}$ are trainable bias parameters of $d$th cross attention.
Then, corresponding cross attention output is computed as
\begin{align}
    \mathbf{A}^{d}_{i,\textrm{ca}} & = \mathrm{Softmax}\left( \frac{\mathbf{Q}^{d}_{i,\textrm{ca}} \left(\mathbf{K}^{d}_{i,\textrm{ca}}\right)^T}{\sqrt{\mathrm{dim}}}\right) \mathbf{V}^{d}_{i,\textrm{ca}}
\end{align}

In the second sublayer of cross attention transformer, fully connected layers with a GELU activation serve as the feed forward network, generating the final output of the $d$th multi-channel attention for the $i$th embedding input, denoted as $\mathbf{X}_i^d$. Residual connections and normalization are applied to each sublayer similar to approach in the self attention transformer. Additionally, multi-channel attention modules have also residual connections between each module.

Following the stacked multi-channel attention module, an additional self attention transformer is employed to compute the attention weights across antenna pairs $(M,N)$. It shares the same structures as the one used in multi-channel attention module. Finally, the network architecture concludes with regular convolutional layers to generate beamforming weights, denoted as $\mathbf{W}_{nn}$.

\subsection{Training Procedure}
The proposed framework offers unsupervised training approach. The objective is to maximize the sum-rate across all UEs. Therefore, the loss function is defined according to the sum-rate maximization problem given in (\ref{sum-rate maximization problem}),
\begin{align} \label{loss_function}
    \mathcal{L}(\bm{\theta};\hat{\mathbf{H}},{\mathbf{W}}_{nn}) = -\sum_{i=1}^N \alpha_i \log(1 + \gamma_i)
\end{align}
where $\bm{\theta}$ denotes the neural network parameters. The loss function is computed by neural network input $\hat{\mathbf{H}}$ and output $f(\bm{\theta}; \hat{\mathbf{H}}) = \mathbf{W}_{nn}$, where $f(\cdot)$ denotes the neural network function. It is important to note that $\gamma_i$ depends on both $\mathbf{H}$ and $\hat{\mathbf{H}}$. Therefore, the performance is also affected by the proposed framework's ability to manage errors in channel estimation.

To compare the performance of the proposed network output $\mathbf{W}_{nn}$, ZFBF weights $\mathbf{W}_{zf}$ and MMSE beamforming weights $\mathbf{\mathbf{W}}_{mmse}$ are considered as baseline techniques. These baseline weights are derived from the channel estimate $\hat{\mathbf{H}}$ and the noise variance $\sigma^2$ as,
\begin{align}\label{zfbf_formula}
    \mathbf{W}_{zf} &= \left(\hat{\mathbf{H}}^H\hat{\mathbf{H}}\right)^{-1}\hat{\mathbf{H}}^H \\
    \mathbf{W}_{mmse} &= \left(\hat{\mathbf{H}}^H\hat{\mathbf{H}}+\sigma^2 \mathbf{I}_N \right)^{-1} \hat{\mathbf{H}}^H 
\end{align}

\section{Experiments}
In our experiments, we evaluate the performance of the proposed framework compared to the baseline techniques ZFBF and MMSE. Spectral efficiency and BLER are considered as performance metrics for throughput improvement and reliable communication. 

\begin{table} [t]
    \begin{center}
    \resizebox{\columnwidth}{!}{%
    \begin{tabular}{| c | c |}
    \hline
    \textbf{Parameter} & \textbf{Value} \\ \hline
    Number of resource blocks (RBs) & 4 (48 subcarriers)  \\ \hline
    Maximum Doppler shift $f_d$ & 260 Hz \\ \hline
    Maximum UE velocity $v$ & 30 m/s \\ \hline
    Carrier frequency $f_c$ & 2.6 GHz \\ \hline
    Subcarrier spacing & 30 kHz \\ \hline
    Transmission time interval (TTI) & 500 $\mu s$ \\ \hline
    Coding rate & $\frac{1}{2}$, $\frac{3}{4}$ \\ \hline
    Modulation scheme & 4QAM, 16QAM, 64QAM, 256QAM \\ \hline
    Training SNR & [-10,20] dB   \\\hline
    Learning rate & $[10^{-5}, 10^{-2}]$ \\ \hline
    $\alpha_{\textrm{la}}$ & 0.5 \\ \hline
    k & 13 \\ \hline
    Minimum training SNR ranges & [15,20],[10,15],[5,10],[0,5],[-10,0] \\ \hline
    \end{tabular}}
    \end{center}
    \caption{System \& training parameters.}
    \label{table:system parameters}
\end{table}

\subsection{System \& Training Specifications}
Experiments are conducted for $2\times8$, $2\times16$ as antenna configurations $N\times M$. Models are trained across a wide SNR range of [-10, 20] dB, covering both low and high SNR regimes in wireless networks. Channel responses are generated using the UMa channel model in the NVIDIA Sionna library \cite{sionna} following 3GPP TR 38.901 specifications \cite{3gppTR38901}. For model training, hyperparameter optimization is performed over optimizers $\{$Adam, AdamW, Radam, RMSprop, Adagrad, Adadelta$\}$ and learning rate schedulers $\{$ReduceLROnPlateau, CosineAnnealing, CosineAnnealingWarmRestarts, ExponentialLR, CyclicLR$\}$ in Optuna \cite{optuna_2019}. Additionally, the Lookahead optimizer is used with a base optimizer for faster convergence and improved stability, with Lookahead steps $k=13$ and update coefficient $\alpha_{\textrm{la}}=0.5$. Specifically, the base optimizer updates model parameters for $k$ steps to obtain fast parameters $\bm{\theta}^{\textrm{fast}}_{t}$, after which they are updated as $\bm{\theta}^{\textrm{slow}}_t = \bm{\theta}^{\textrm{slow}}_{t-k} + \alpha_{\textrm{la}}(\bm{\theta}^{\textrm{fast}}_{t} - \bm{\theta}^{\textrm{slow}}_{t-k})$. A curriculum learning approach is applied, where training progresses from simpler to more complex tasks by adjusting the SNR at each stage. The maximum SNR is fixed at 20 dB, while the minimum SNR value for each stage is optimized through hyperparameter tuning. System and training parameters are summarized in Table~\ref{table:system parameters}.

\subsection{Results and Analysis}
\begin{figure}[t]
    \begin{center}
     	\subfigure[]{%
     	\includegraphics[width=0.48\linewidth]{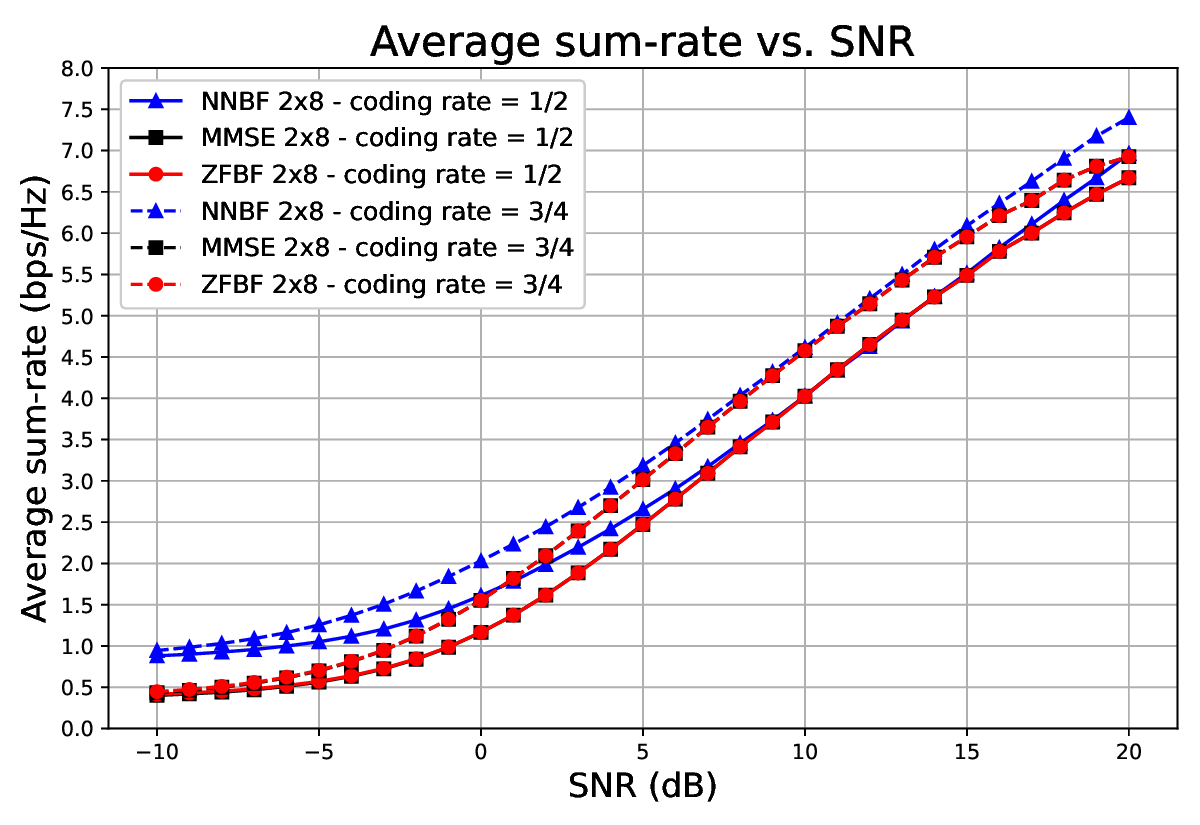}}
     	\subfigure[]{%
     	\includegraphics[width=0.48\linewidth]{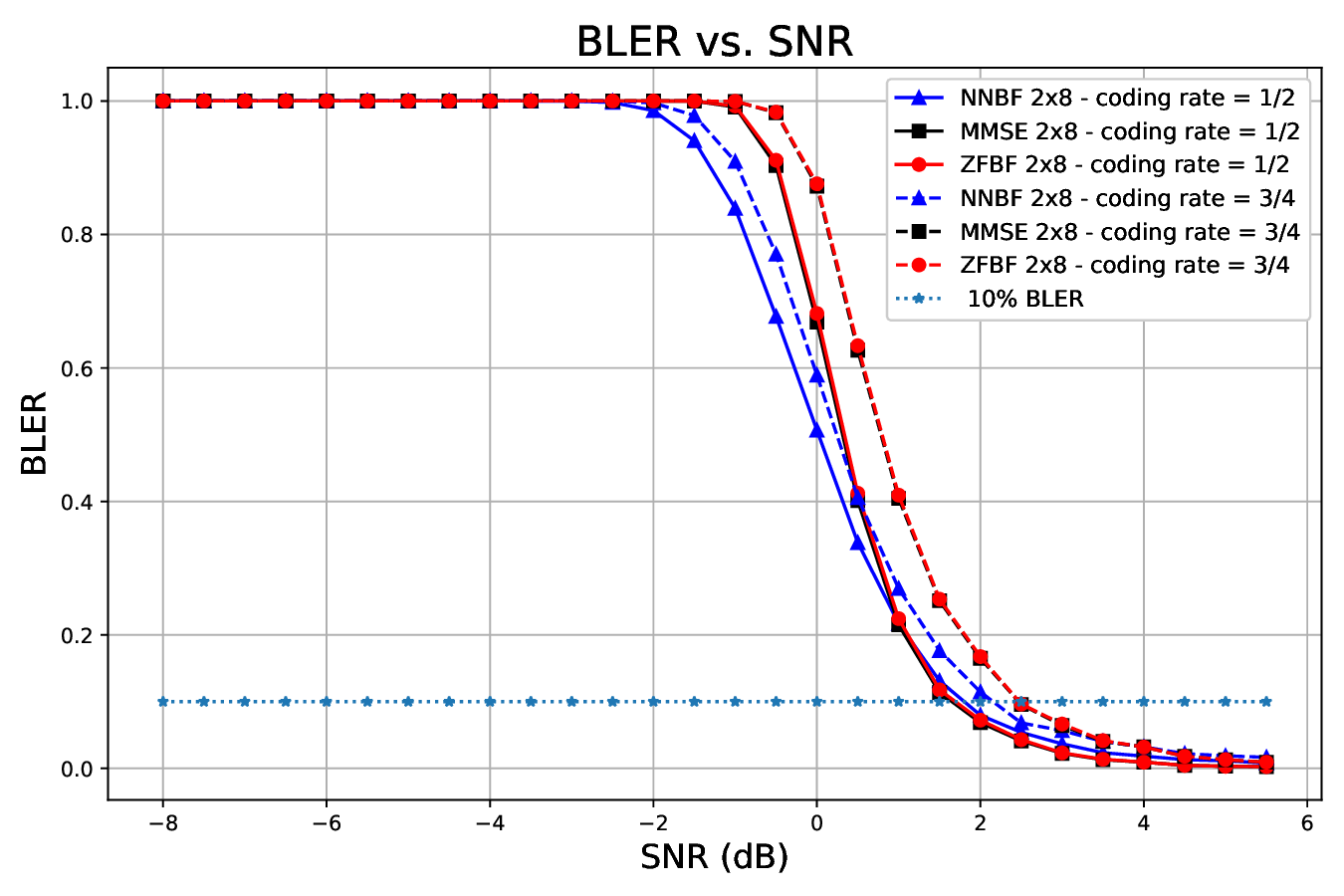}}
    \end{center}
     \caption{Performance comparison of NNBF with baseline methods ZFBF and MMSE when the coding rate can be $\frac{1}{2}$ and $\frac{3}{4}$. The results are shown for a system with $M=8$, $N=2$, 4QAM modulation, and for stationary UEs.} 
    \label{4qam 2x8}
\end{figure}

Fig.~\ref{4qam 2x8} illustrates the effect of coding rate, i.e., the ratio of useful bits to the total number of bits transmitted including redundancy. The system setup is presented for $2\times8$ scenario while the UEs are stationary and modulation type is 4QAM. The coding rate can be $\frac{1}{2}$ or $\frac{3}{4}$. In Fig.~\ref{4qam 2x8}(a), for average sum-rate metric, it can be seen that the proposed framework NNBF outperforms the baseline techniques in the SNR range $[-10,5]$ dB, which includes both low and high SNR regimes for practical wireless network scenarios. Moreover, NNBF with coding rate of $\frac{1}{2}$ (solid blue triangle line) performs better than the baseline techniques with coding rate of $\frac{3}{4}$ (dashed black square and dashed circle lines) for low SNR regime $[-10,-3]$ dB although higher coding rate is expected to provide higher throughput. This result shows that the proposed framework could achieve higher throughput even with less coding rate. In addition to spectral efficiency, we can compare the proposed framework with baseline techniques in terms of communication reliability. Fig.~\ref{4qam 2x8}(b) exhibits the BLER performances. NNBF has lower BLER than MMSE and ZFBF while they can achieve $10\%$ BLER around comparable channel SNR values, $[1.5,2.5]$ dB. Therefore, it shows that the proposed framework provides higher spectral efficiency without sacrificing the reliability of communication.

\begin{figure}[h]
    \begin{center}
     	\subfigure[]{%
     	\includegraphics[width=0.48\linewidth]{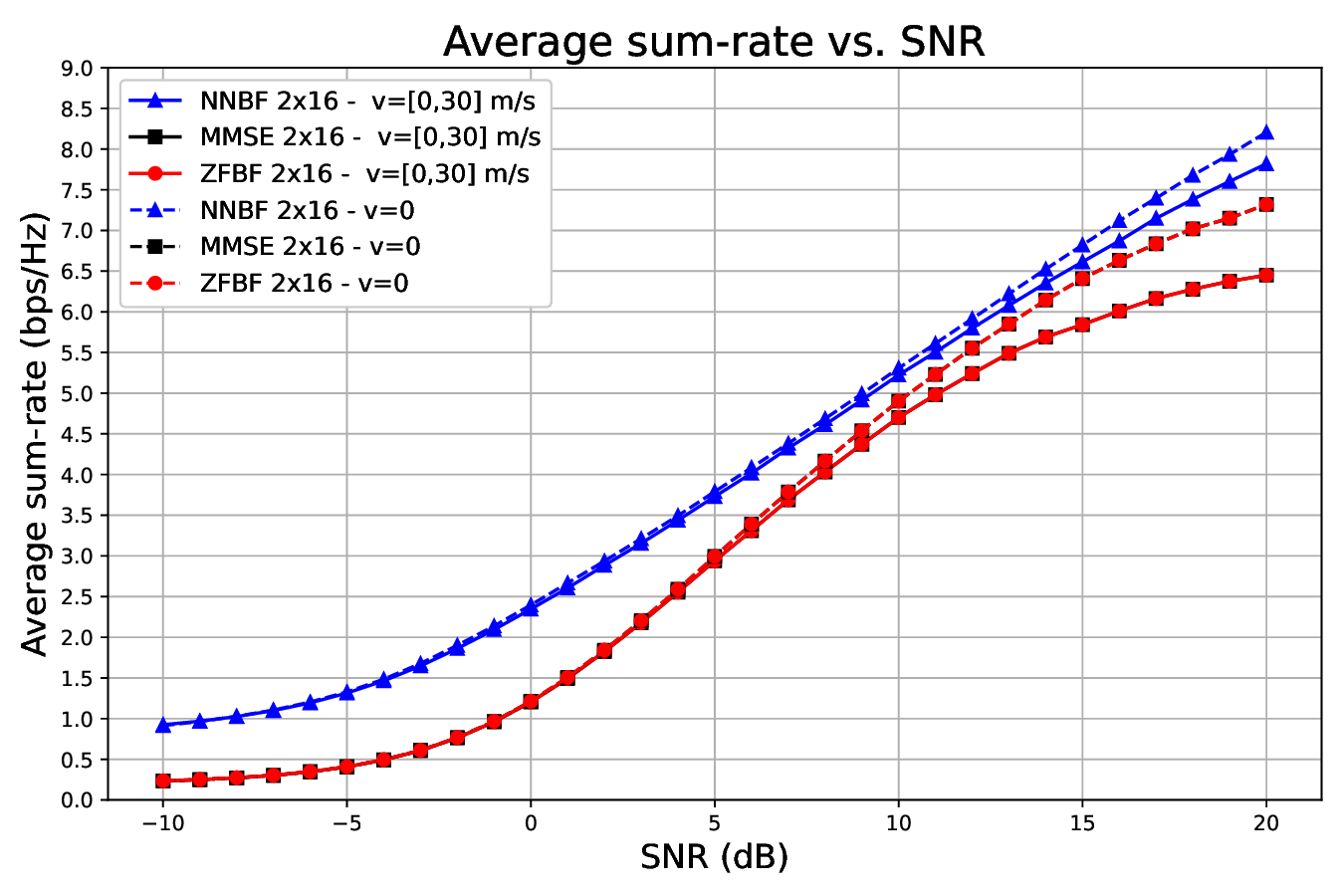}}
     	\subfigure[]{%
     	\includegraphics[width=0.48\linewidth]{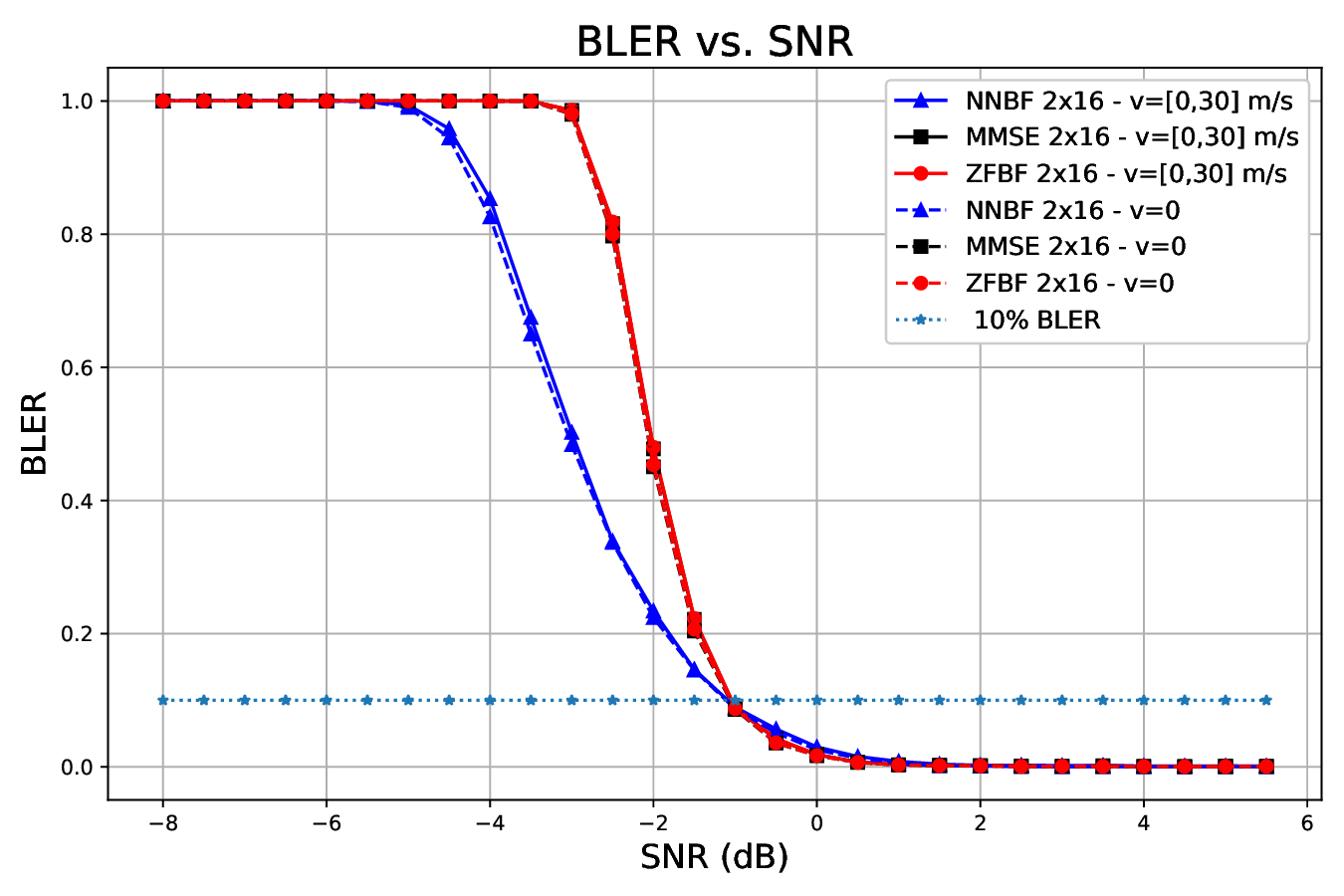}}
    \end{center}
     \caption{Performance comparison of NNBF with baseline methods ZFBF and MMSE for stationary and mobile UEs with a maximum velocity of 30 m/s. The results are shown for a system with $M=16$, $N=2$, 4QAM, and coding rate of $\frac{1}{2}$.} 
    \label{4qam 2x16 velocity vs stationary}
\end{figure}

Fig.~\ref{4qam 2x16 velocity vs stationary} exhibits the performance comparison of NNBF with ZFBF and MMSE beamforming for stationary and mobile UEs. Mobile UEs can have velocity up to $30$ m/s. The system configuration is demonstrated for $2\times 16$ setup with 4QAM modulation and the coding rate is set to $\frac{1}{2}$. In the context of mobile communications, Doppler shift frequency $f_d=\frac{f_c v \cos{\phi}}{c}$ describes the frequency shift experienced by the signal due to the motion of the mobile UE, where $c$ is the speed of light, $v$ is the velocity of UE and $\phi$ denotes the angle between the BS and UE. $f_c$ is the carrier frequency and is set to $2.6 $ GHz in our experiments. Then, the maximum Doppler shift for experiments presented in Fig.~\ref{4qam 2x16 velocity vs stationary} is $260$ Hz, considering that maximum velocity is $30$ m/s. Fig.~\ref{4qam 2x16 velocity vs stationary} shows that the proposed framework NNBF can tolerate the effect of UE mobility, which is inevitable for urban network scenarios. Specifically in term of spectral efficiency, we observe larger drop in average sum-rate after $10$ dB for baseline techniques while NNBF has only negligible decrease.

\begin{figure*}[ht!]
    \begin{center}
     	\subfigure[]{%
     	\includegraphics[width=0.24\linewidth]{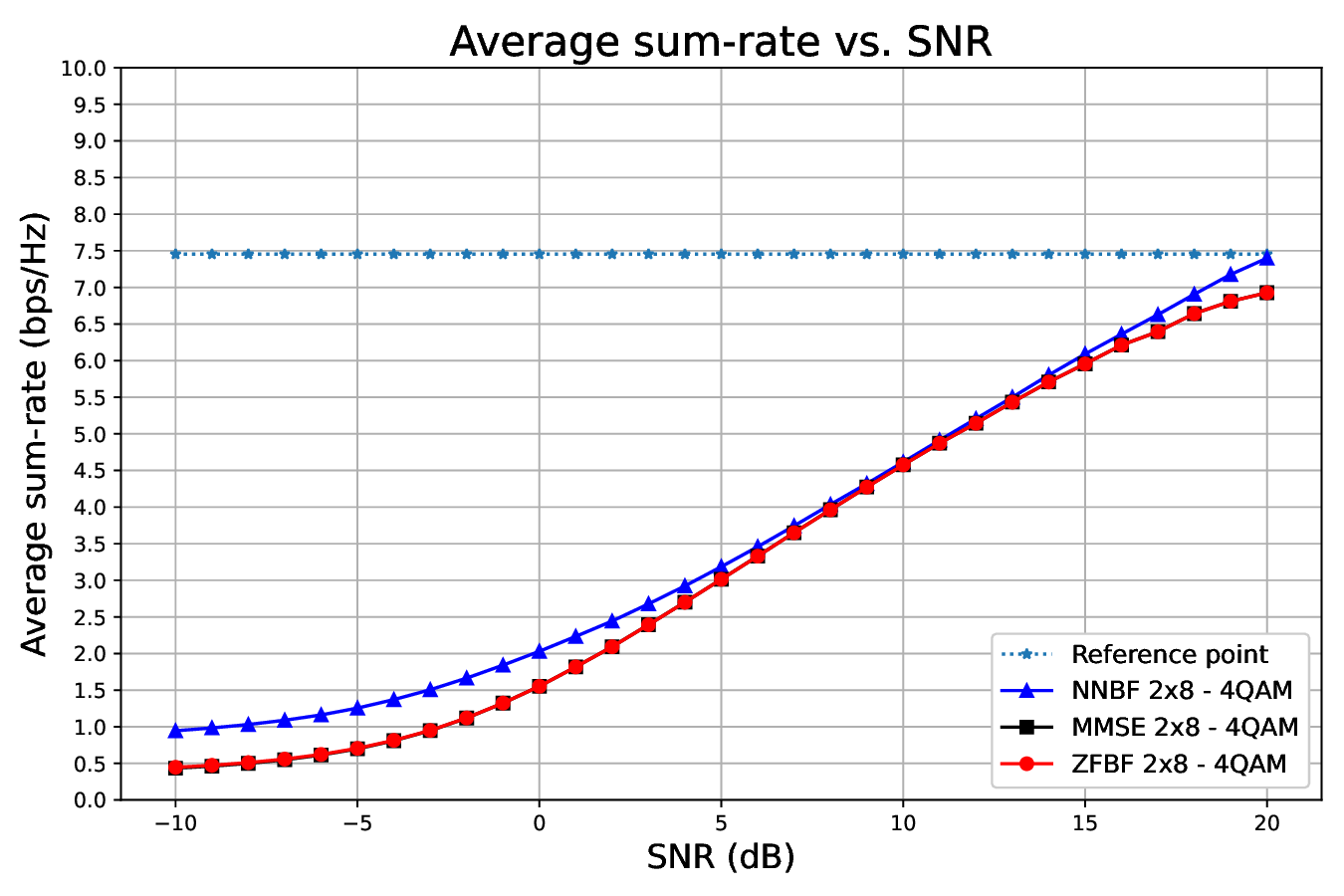}}
     	\subfigure[]{%
     	\includegraphics[width=0.24\linewidth]{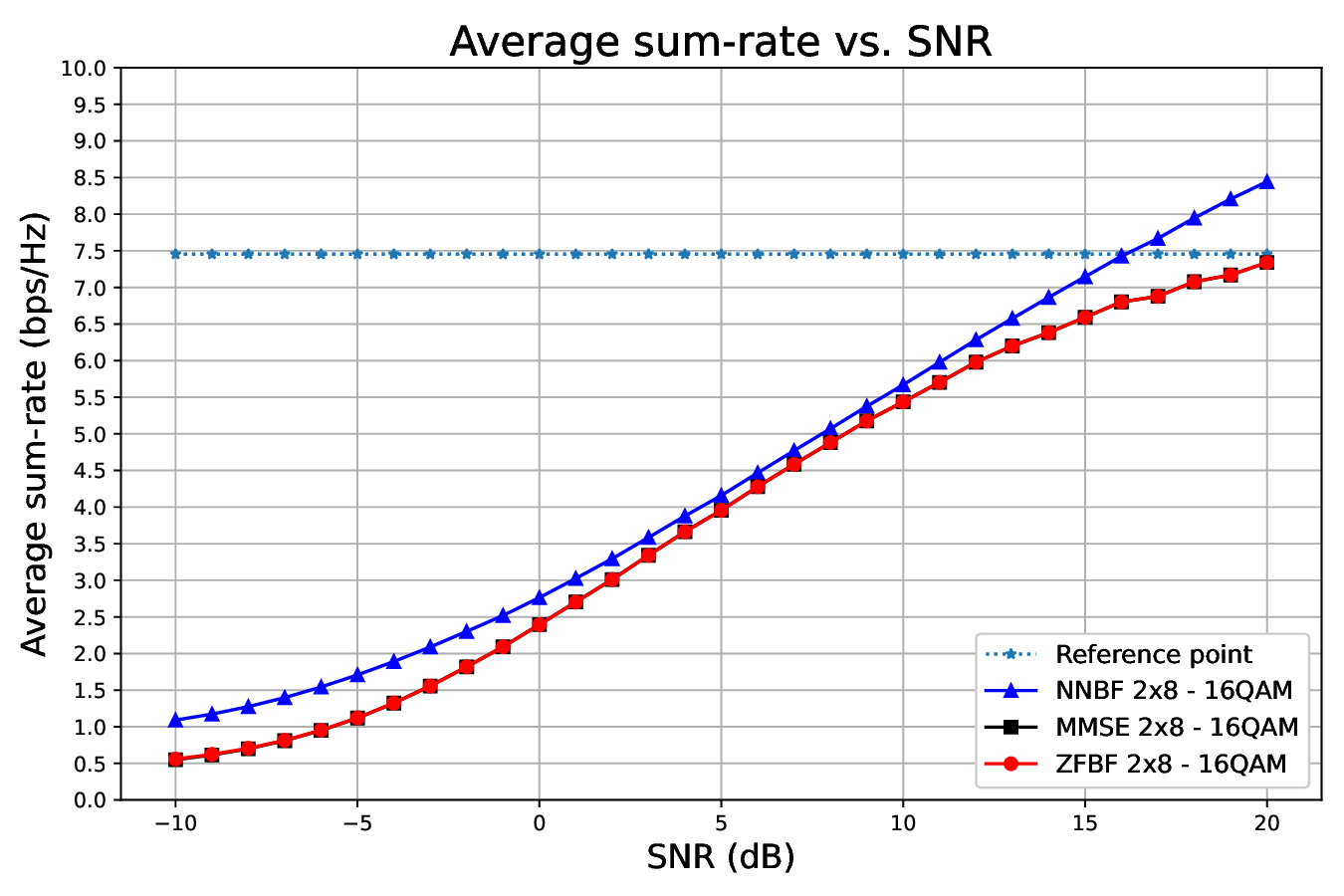}}
     	\subfigure[]{%
     	\includegraphics[width=0.24\linewidth]{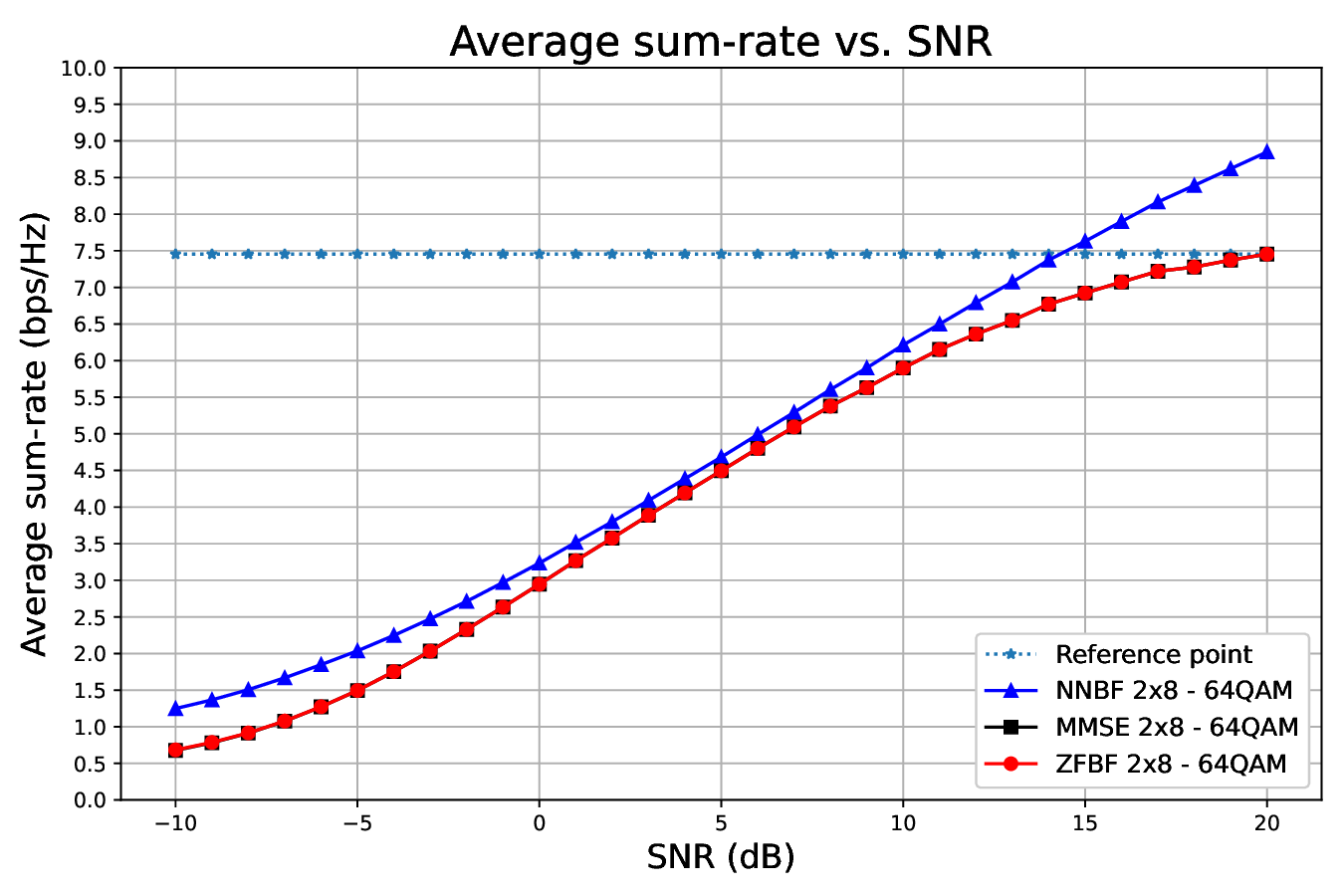}}
            \subfigure[]{%
     	\includegraphics[width=0.24\linewidth]{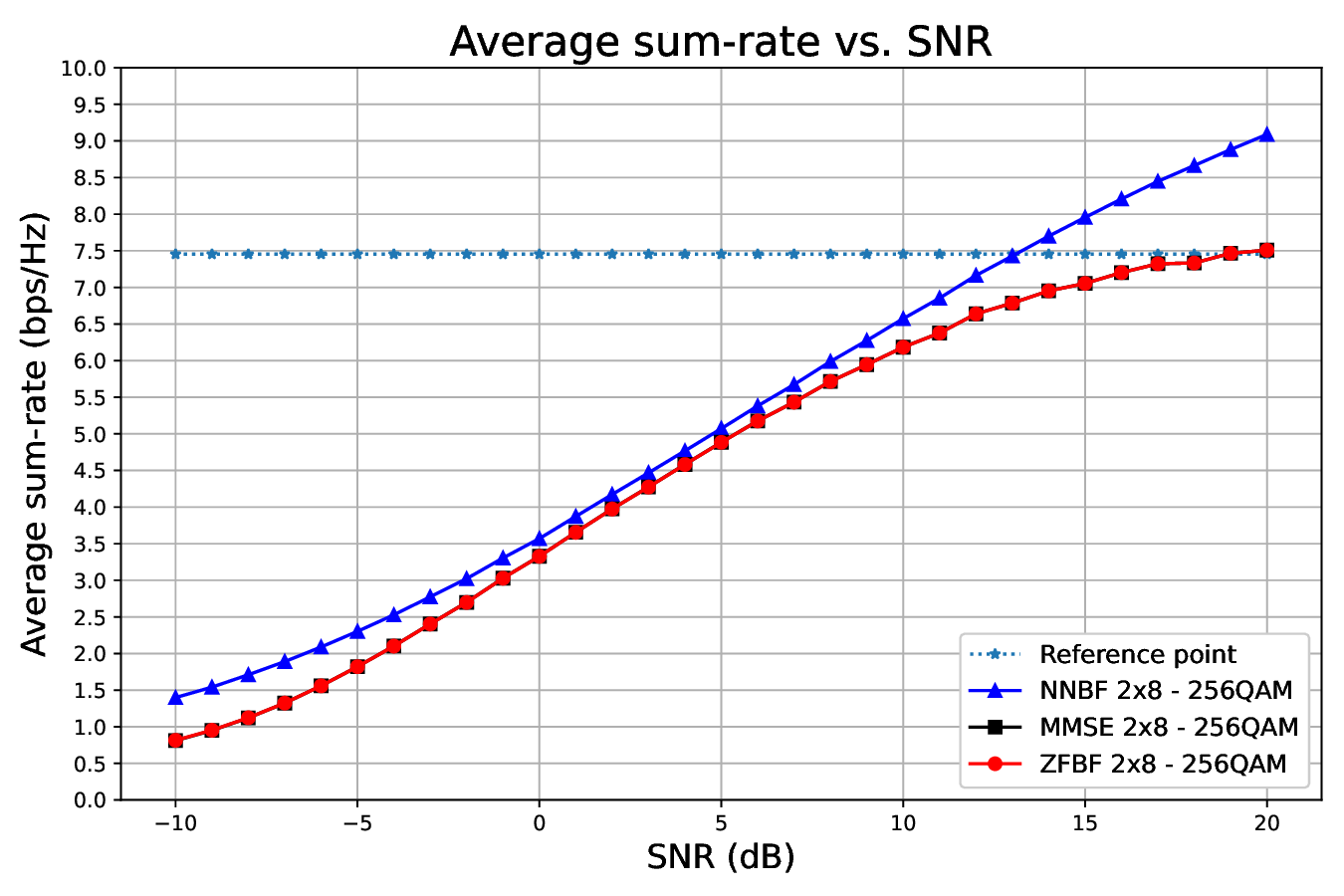}}
    \end{center}
    \caption{Average sum-rate performance comparison of NNBF with baseline methods ZFBF and MMSE for increasing order of modulations: 4QAM, 16QAM, 64QAM, 256QAM. The results are shown for a system with $M=8$, $N=2$, coding rate of $\frac{3}{4}$, and stationary UEs.}
    \label{higher order modulation sinr}
\end{figure*}

\begin{figure}[h!]
    \begin{center}
     	\includegraphics[width=1\linewidth]{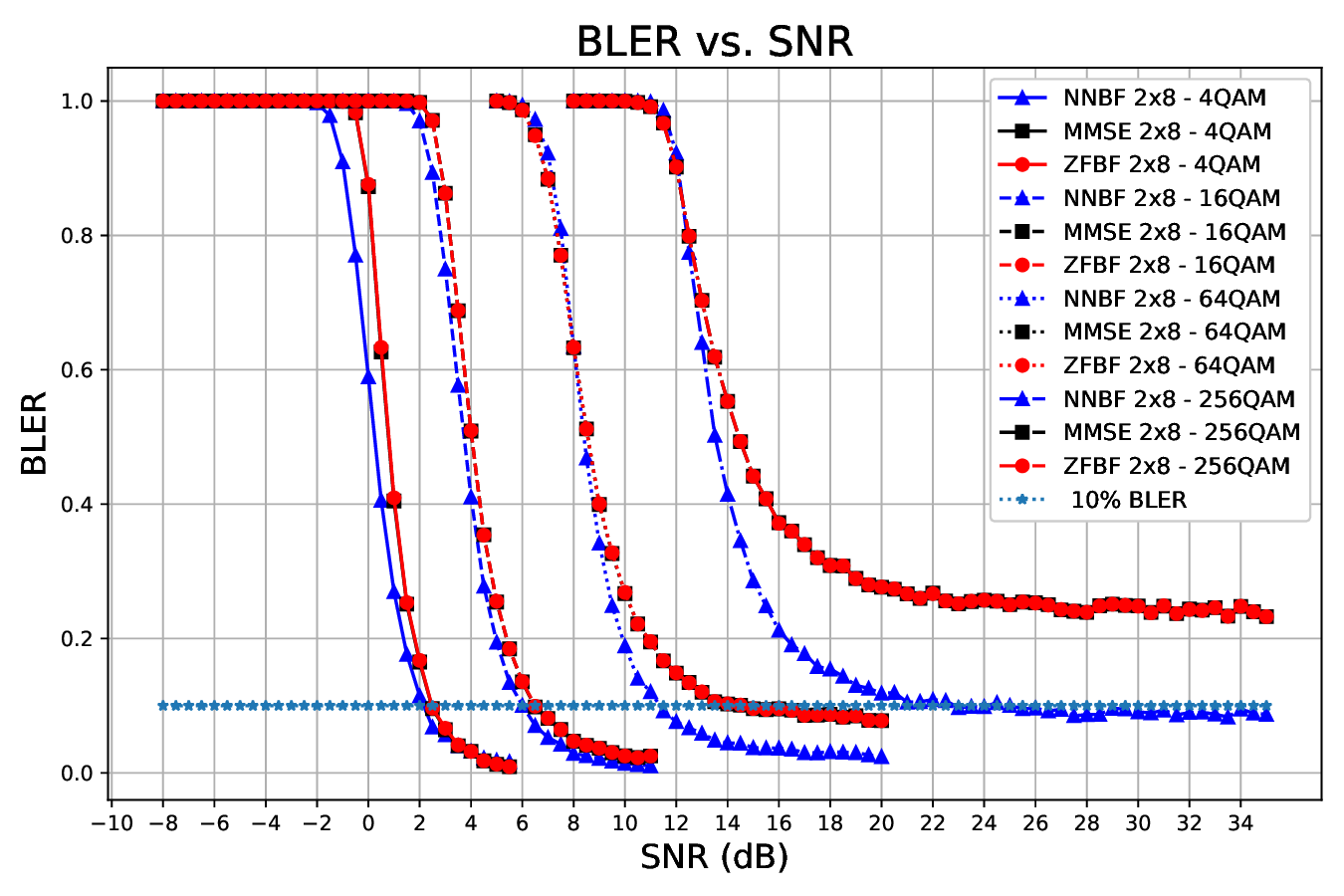}
    \end{center}
     \caption{BLER performance comparison of NNBF with baseline methods ZFBF and MMSE for increasing order of modulations: 4QAM, 16QAM, 64QAM, 256QAM. The results are shown for a system with $M=8$, $N=2$, coding rate of $\frac{3}{4}$, and stationary UEs.}
    \label{higher order modulation bler}
\end{figure}

Finally, Fig.~\ref{higher order modulation sinr} and Fig.~\ref{higher order modulation bler} exhibit the average sum-rate and BLER performances of the proposed framework compared to baseline techniques for increasing orders of modulation, respectively. In Fig.~\ref{higher order modulation sinr}, it can be seen that average sum-rate for baseline techniques converges to the reference point as the modulation order increases. However, average sum-rate for NNBF further improves with the increase of modulation order, suggesting that it may eventually converge to a higher value. It demonstrates that the proposed NNBF model can better exploit the additional spectral efficiency offered by higher order modulation schemes while the baseline techniques are limited by their sub-optimal beamforming strategies. Furthermore, in all previous experiment results, similar BLER performances are achieved for NNBF and baseline techniques when modulation type is taken to be 4QAM. However, as the order of the modulation increases, the superiority of NNBF BLER performance can be seen over MMSE and ZFBF performances in Fig~\ref{higher order modulation bler}. For 64QAM, NNBF could achieve $10\%$ BLER at $11.5$ dB while MMSE and ZFBF can achieve  $10\%$ BLER at $13.5$ dB. For 256QAM, MMSE and ZFBF converges to $20\%$ BLER while NNBF still can go below $10\%$ BLER.

\section{Conclusion}
We introduced NNBF,  a novel unsupervised deep learning framework that combines depthwise separable convolutions and transformers to generate beamforming weights under imperfect CSI for a MU-SIMO system in dense urban environments. NNBF improves throughput by sum-rate maximization while maintaining reliable communication. Experimental results demonstrates that NNBF outperforms ZFBF and MMSE beamforming across a broad SNR range, covering both low and high SNR regimes. Notably, NNBF achieves significant performance gains with higher order modulation schemes and surpasses baseline methods across different coding rate and mobility conditions.

\bibliographystyle{unsrt}
\bibliography{main}

\begin{thebibliography}{10}

\bibitem{erpek2020deep}
T.~Erpek, T.~J. O'Shea, Y.~E. Sagduyu, Y.~Shi, and T.~C. Clancy.
\newblock Deep learning for wireless communications.
\newblock 2020.
\newblock Available online at arXiv:2005.06068.

\bibitem{2011iterativeWeightedMMSE}
Q.~Shi, M.~Razaviyayn, Z.~Luo, and C.~He.
\newblock An iteratively weighted {MMSE} approach to distributed sum-utility maximization for a {MIMO} interfering broadcast channel.
\newblock {\em IEEE Transactions on Signal Processing}, 59(9):4331--4340, September 2011.

\bibitem{2008_WSR_WMMSE}
S.~Christensen, R.~Agarwal, E.~de~Carvalho, and J.M. Cioffi.
\newblock Weighted sum-rate maximization using weighted {MMSE} for {MIMO-BC} beamforming design.
\newblock {\em IEEE Transactions on Wireless Communications}, 7(12):4792--4799, December 2008.

\bibitem{ClancyCognitive}
C.~Clancy, J.~Hecker, E.~Stuntebeck, and T.~O'Shea.
\newblock Applications of machine learning to cognitive radio networks.
\newblock {\em IEEE Wireless Communications}, 14(4):47--52, August 2007.

\bibitem{AppML}
Y.~Sun, M.~Peng, Y.~Zhou, Y.~Huang, and S.~Mao.
\newblock Application of machine learning in wireless networks: Key techniques and open issues.
\newblock {\em IEEE Communications Surveys \& Tutorials}, 21(4):3072--3108, June 2019.

\bibitem{Sun2018}
H.~Sun, X.~Chen, Q.~Shi, M.~Hong, X.~Fu, and N.~D. Sidiropoulos.
\newblock Learning to optimize: Training deep neural networks for interference management.
\newblock {\em IEEE Transactions on Signal Processing}, 66(20):5438--5453, October 2018.

\bibitem{Guo2018NetworkDF}
J.~Guo, Y.~Li, W.~Lin, Y.~Chen, and J.~Li.
\newblock Network decoupling: from regular to depthwise separable convolutions.
\newblock In {\em BMVC}, 2018.

\bibitem{Naeem2021}
M.~Naeem, G.~D. Pietro, and A.~Coronato.
\newblock Application of reinforcement learning and deep learning in multiple-input and multiple-output {MIMO} systems.
\newblock {\em Sensors}, 22:309, 2021.

\bibitem{DT_transformer}
J.~Zhang, J.~Li, Z.~Wang, L.~Shi, S.~Jin, W.~Chen, and V.~H. Poor.
\newblock Decision transformers for wireless communications: A new paradigm of resource management.
\newblock {\em IEEE Wireless Communications}, pages 1--7, 2025.

\bibitem{atteion_based_nn}
D.~Luan and J.~Thompson.
\newblock Attention based neural networks for wireless channel estimation.
\newblock In {\em IEEE VTC}, June 2022.

\bibitem{3gppTR38901}
{3GPP}.
\newblock {Study on channel model for frequencies from 0.5 to 100 GHz}.
\newblock Technical Report TR 38.901, {3rd Generation Partnership Project (3GPP)}, April 2022.
\newblock Version 17.0.0.

\bibitem{vahapoglu2023deep}
C.~Vahapoglu, T.~J. O'Shea, T.~Roy, and S.~Ulukus.
\newblock Deep learning based uplink multi-user {SIMO} beamforming design.
\newblock In {\em IEEE ICMLCN}, May 2023.

\bibitem{vahapoglu2024_joint_power_bf}
C.~Vahapoglu, T.~J. O’Shea, T.~Roy, and S.~Ulukus.
\newblock Deep learning based joint multi-user {MISO} power allocation and beamforming design.
\newblock In {\em IEEE MILCOM}, October 2024.

\bibitem{Chollet2016XceptionDL}
F.~Chollet.
\newblock Xception: Deep learning with depthwise separable convolutions.
\newblock In {\em IEEE CVPR}, July 2017.

\bibitem{multi-channel-transformer-sr}
F.~Chang, M.~Radfar, A.~Mouchtaris, B.~King, and S.~Kunzmann.
\newblock End-to-end multi-channel transformer for speech recognition.
\newblock In {\em ICASSP}, 2021.

\bibitem{Vaswani2017AttentionIA}
A.~Vaswani, N.~Shazeer, N.~Parmar, J.~Uszkoreit, L.~Jones, A.~Gomez, L.~Kaiser, and I.~Polosukhin.
\newblock Attention is all you need.
\newblock In {\em Neural Information Processing Systems}, 2017.

\bibitem{sionna}
J.~Hoydis, S.~Cammerer, F.~{Ait Aoudia}, A.~Vem, N.~Binder, G.~Marcus, and A.~Keller.
\newblock Sionna: An open-source library for next-generation physical layer research.
\newblock {\em arXiv preprint}, March 2022.

\bibitem{optuna_2019}
T.~Akiba, S.~Sano, T.~Yanase, T.~Ohta, and M.~Koyama.
\newblock Optuna: A next-generation hyperparameter optimization framework.
\newblock In {\em KDD}, 2019.

\end{thebibliography}
\end{document}